\documentclass[aps,preprint]{revtex4}%
\usepackage{amsfonts}
\usepackage{amsmath}
\usepackage{amssymb}
\usepackage{graphicx}%
\begin{document}
\title{Entanglement \ of dipolar coupling spins}
\author{G.B. Furman$^{1,2}$, V. M. Meerovich$^{1}$, and V.L. Sokolovsky$^{1}$ }
\affiliation{$^{1}$Department of Physics, Ben-Gurion University of the Negev, POB 653,
Beer-Sheva 84105, Israel}
\affiliation{$^{2}$Ohalo College, POB 222, Qazrin, 12900, Israel}

\begin{abstract}
Entanglement of dipole-dipole interacting spins 1/2 is usually investigated
when the energy of interaction with an external magnetic field (the Zeeman
energy) is greater than the energy of dipole interactions by three orders.
Under this condition only a non-equilibrium state of the spin system, realized
by pulse radiofrequence irradiations, results in entanglement.

The present paper deals with the opposite case: the dipolar interaction energy
is the order of magnitude or even larger than the Zeeman one. It was shown
that entanglement appears under the thermodynamic equilibrium conditions and
the concurrence reaches the maximum when the external field is directed
perpendicular to the vector connecting the nuclei. For this direction of the
field and a system of two spins with the Hamiltonian accounting the realistic
dipole-dipole interactions in low external magnetic field , the exact
analytical expression for concurrence was also obtained. The condition of the
entanglement appearance and the dependence of concurrence on the external
magnetic field, temperature, and dipolar coupling constant were studied.

\end{abstract}
\startpage{1}
\maketitle

\section{Introduction}

Appreciation of the role of quantum entanglement \cite{Benenti books
2007,Amico 2008 rev,HorodeckiR 2009 rev} as a resource in quantum
teleportation \cite{Bennett CH 1993}, quantum communication \cite{Bennett CH
1984}, quantum computation \cite{Shor P 1994}, and quantum metrology
\cite{Cappellaro P 2005,Roos2006} has stimulated intensive qualitative and
quantitative research. Entanglement, as the quantum correlation, can bring up
richer possibilities in the various fields of modern technology. Therefore, in
the past few years great efforts have been done to understand and create
entanglement. Entanglement between two quantum systems can be generated due to
their interaction only \cite{Benenti books 2007,Amico 2008 rev,HorodeckiR 2009
rev,Bose S 2007}. It has recently been shown that, in a chain of nuclear spins
$s$ = 1/2, which is described by the idealized XY model for a spin system
under the thermodynamic equilibrium conditions, entanglement appears at very
low temperatures $T\approx0.5$
%TCIMACRO{\U{b5}}%
%BeginExpansion
$\mu$%
%EndExpansion
K \cite{Fel'dman2007}.

In most real quantum systems, such as dipolar coupling spin system, specific
conditions for creation of the entangled states are requested. In two-and
three-spin \cite{Doronin2003} and many-spin \cite{Fel'dman2008} clusters of
protons subjected to a strong magnetic field, truncated dipole-dipole
interactions and multiple pulse radiofrequence irradiations, the entangled
state of a spin pair emerges at temperatures $T\thickapprox20$ mK. In these
papers the cases were considered where the energy of interaction of the spins
with the external magnetic field (the Zeeman energy) is greater than the
energy of dipole interactions by three orders \cite{Doronin2003,Fel'dman2008}.
It was shown that at this condition only a non-equilibrium state of the spin
system, realized by pulse radiofrequence irradiations, results in entanglement
\cite{Fel'dman2008,Furman2008}.

The present paper deals with the case opposite to those considered previously
\cite{Doronin2003,Fel'dman2008}: the dipolar interaction energy is the order
of magnitude or even greater than the Zeeman one. We investigate entanglement
of two spins coupled by the realistic dipole-dipole interactions in a low
external magnetic field under the thermodynamic equilibrium conditions. We
study dependence of the critical temperature and magnetic field at which
entanglement appears in this system on a dipolar coupling constant.

\section{Hamiltonian of dipolar coupling spin system and \textbf{concurrence
between nuclear spin}s\textbf{ 1/2}}

Let us consider a system of $N$ spins coupled by long-range dipolar
interactions and subjected to an external magnetic field, $\vec{H}_{0}%
=H_{0}\vec{z}$. The total Hamiltonian of this interacting system can be
written as
\begin{equation}
H=H_{z}+H_{dd} \tag{1}%
\end{equation}
where the Hamiltonian $H_{z}$ describes the Zeeman interaction between the
nuclear spins and external magnetic field (here we used $\hbar=1$)%

\begin{equation}
H_{z}=\omega_{0}\sum_{k=1}^{N}I_{k}^{z}, \tag{2}%
\end{equation}
$\omega_{0}=\gamma H_{0}$ is the energy difference between the excited and
ground states of an isolated spin, $\gamma$ is the gyromagnetic ratio of a
spin, $I_{k}^{z}$ is the projection of the angular spin momentum operator on
the $z$- axes. The Hamiltonian $H_{dd}$ describing dipolar interactions in an
external magnetic field \cite{Abragam 1982}:
\begin{align}
H_{dd}  &  =\sum_{j<k}\frac{\gamma^{2}}{r_{jk}^{3}}\{\left(  1-3\cos^{2}%
\theta_{jk}\right)  \left[  I_{j}^{z}I_{k}^{z}-\frac{1}{4}\left(  I_{j}%
^{+}I_{k}^{-}+I_{j}^{-}I_{k}^{+}\right)  \right]  -\nonumber\\
&  \frac{3}{4}\sin2\theta_{jk}\left[  e^{-i\varphi_{jk}}\left(  I_{j}^{z}%
I_{k}^{+}+I_{j}^{+}I_{k}^{z}\right)  +e^{i\varphi_{jk}}\left(  I_{j}^{z}%
I_{k}^{-}+I_{j}^{-}I_{k}^{z}\right)  \right]  -\frac{3}{4}\sin^{2}\theta
_{jk}\left[  e^{-2i\varphi_{jk}}I_{j}^{+}I_{k}^{+}+e^{2i\varphi_{jk}}I_{j}%
^{-}I_{k}^{-}\right]  \} \tag{3}%
\end{align}
where $r_{jk}$, $\theta_{jk}$, and $\varphi_{jk}$ are the spherical
coordinates of the vector $\vec{r}_{jk}$ connecting the $j-$th and $k-$th
nuclei in a coordinate system with the $z$-axis along the external magnetic
field, $\vec{H}_{0}$, $I_{j}^{+}$and $I_{j}^{-}$ are the raising and lowering
spin angular momentum operators of the $j$-th spin. We consider the situation
when it is necessary to take into account all the terms of the Hamiltonian of
the dipole-dipole interactions, and not trusnckete any ones.

In the thermodynamic equilibrium the considered system is described by the
density matrix
\begin{equation}
\rho=Z^{-1}\exp\left(  -\frac{H}{k_{B}T}\right)  , \tag{4}%
\end{equation}
where $Z=Tr\left\{  \exp\left(  -H/k_{B}T\right)  \right\}  $ is the partition
function, $k_{B}$ is the Boltzamnn constant, and $T$ is the temperature. We
will analyze entanglement in the spin system described by the density matrix (4).

In order to quantify entanglement, the concurrence $C$ is usually used
\cite{Wootters1998}. For the maximally entangled states, the concurrence is
$C=1$, while for the separable states $C=0$. The concurrence between the
quantum states of two spins presented in the Hilbert space as a matrix
$4\times4$ is expressed by the formula \cite{Wootters1998}
\begin{equation}
C=\max\left\{  0,2\lambda-\sum_{i=1}^{4}\lambda_{i}\right\}  \tag{5}%
\end{equation}
where $\lambda=\max\left\{  \lambda_{1},\lambda_{2},\lambda_{3},\lambda
_{4}\right\}  $ and $\lambda_{i}$ $\left(  i=1,2,3,4\right)  $ are the square
roots of the eigenvalues of the product
\begin{equation}
R=\rho\tilde{\rho} \tag{6}%
\end{equation}
with%
\begin{equation}
\tilde{\rho}=\left(  \sigma_{y}\otimes\sigma_{y}\right)  \bar{\rho}\left(
\sigma_{y}\otimes\sigma_{y}\right)  \tag{7}%
\end{equation}
where $\bar{\rho}$ the complex conjugation of the density matrix (4) and
$\sigma_{y}$ is the Pauli matrix
\begin{equation}
\sigma_{y}=%
\begin{pmatrix}
0 & -i\\
i & 0
\end{pmatrix}
. \tag{8}%
\end{equation}

\section{\textbf{Entanglement in pair of spins}}

We examine dependence of the concurrence, $C$, between states of the two spins
1/2 on the magnetic field strength and its direction, dipolar coupling
constant, and temperature. The numerical calculation of \emph{ }entanglement
of the spins at arbitrary orientation of the magnetic field are performed
using the software based on the Mathematica package. The results of the
numerical calculation show that concurrence reaches its maximum at the case of
$\theta=\frac{\pi}{2}\ $and $\varphi=0$ (Fig. 1) and we will consider this
case below. This orientation of the spins allows us to obtain for concurrence
as an exact analytical function of the temperature, magnetic field and dipolar
coupling constant $\gamma^{2}/r_{12}^{3}$. Using the exact diagonalization of
the density matrix (4) we obtain the concurrence in the following form:
\begin{equation}
C\left(  \beta,d\right)  =\max\left\{  0,\frac{A_{+}-A_{-}-e^{\frac{d}{2}%
}\cosh\frac{d}{4}}{\left(  e^{\frac{d}{2}}\cosh\frac{d}{4}+\cosh\frac
{\sqrt{16\beta^{2}+9d^{2}}}{4}\right)  }\right\}  , \tag{9}%
\end{equation}
where%
\begin{equation}
A_{\pm}=\frac{1}{2}\sqrt{\frac{16\beta^{2}+9d^{2}\cosh\frac{\sqrt{16\beta
^{2}+9d^{2}}}{2}\pm6d\sinh\frac{\sqrt{16\beta^{2}+9d^{2}}}{2}\sqrt{16\beta
^{2}+9d^{2}\cosh^{2}\frac{\sqrt{16\beta^{2}+9d^{2}}}{4}}}{16\beta^{2}+9d^{2}}%
}, \tag{10}%
\end{equation}
$\allowbreak$with $\beta=\frac{\omega_{0}}{k_{B}T}$ and $d=\frac{\gamma^{2}%
}{r_{12}^{3}k_{B}T}.$

At high temperature and low magnetic field $\left(  \beta<<1\right)  $ and/or
small dipolar coupling constant ($d<<1)$ the expression in the figure brackets
(9) becomes negative and, therefore, entanglement is zero. Equating this
expression to zero we obtain the critical parameters: temperature $T_{c}$,
strength of magnetic field $H_{c}$, and dipolar coupling constant at which the
entanglement appears in a spin pair. \emph{ }Figure 2 presents the phase
diagram in which the boundary between the entangled and separated states is
determined from equation:%

\begin{equation}
A_{+}-A_{-}-e^{\frac{d}{2}}\cosh\frac{d}{4}=0. \tag{11}%
\end{equation}
For example, at $d=1$ entanglement can be achieved at $\beta>2.26$. To
\ estimate the critical temperature let us consider fluorine with
$\gamma=\allowbreak4.\,\allowbreak0025\frac{kHz}{G}$ and the dipolar
interaction energy typically of order of a few kHz (in frequency units)
\cite{Abragam 1982}. Taking $H_{0}=$ $3$ G we have $\omega_{0}=12$ kHz, which
leads to $T_{c}=0.33$
%TCIMACRO{\U{3bc}}%
%BeginExpansion
$\mu$%
%EndExpansion
K. The estimated value of temperature is in good agreement with those reported
early for the spin system $s=1/2$ with the $XY$ Hamiltonian in absence of a
magnetic field \cite{Fel'dman2007}. Both models give also qualitatively
similar dependences of concurrence on temperature but the model considered by
us predicts appearance of entanglement in an external magnetic field higher
than the critical value. It is interesting that the ordered states, such as
antiferromagnetic, of nuclear spins were observed in a calcium-fluoride
$CaF_{2}$ single crystal at $T=$ $0.34$
%TCIMACRO{\U{3bc}}%
%BeginExpansion
$\mu$%
%EndExpansion
K\ \cite{Abragam 1982,Abragam 1974}. This structure is characterized by
domains in the form of layers perpendicular to the external magnetic field.
The magnetization inside the crystal is parallel to the external field and
reverses its sign from a layer to a layer, while the total magnetization is
zero. It is well know that when the magnetization reaches the maximum, all the
spins are aligned along the field and entanglement is absent \emph{
}\cite{Amico 2008 rev,HorodeckiR 2009 rev}. Entanglement can appear if
magnetization is less than its maximum. Therefore, it is reasonable to assume
that simultaneously with the transition to the ordered state there arises
entanglement of spins from different layers.

Figure 3 shows concurrence as a function of both parameters $\beta$ and $d$ at
$\theta=\frac{\pi}{2}\ $and $\varphi=0$. At large temperature and low magnetic
field concurrence is zero. One can see that the concurrence increases with the
magnetic field and inverse temperature and reaches its maximum. Then the
concurrence decreases. \ Figure 4 shows the concurrence as a function of the
magnetic field at a constant temperature, (Fig. 4a) and as a function of the
inverse temperature at a constant magnetic field\emph{ } (Fig. 4b). In the
both cases concurrence remains zero up to a certain value of the magnetic
field (Fig. 4a) or of the inverse temperature (Fig. 4b), which depends on the
coupling constant. The following increase of the magnetic field or inverse
temperature leads to eventually rising. In the case of increasing the magnetic
field concurrence increases up to the maximum \ and then decreases as a
magnetic field \emph{ }increases (Fig.4a). Another behavior is observed at an
increase of the inverse temperature (Fig. 4b): concurrence monotone grows with
$1/T$ \ up to a\emph{ }steady\emph{ }value depending on a magnetic field and
the following increase of the inverse temperature does not cause any change of concurrence.

\section{\textbf{Discussion and conclusions}}

It was obtained that in zero magnetic field the system is in a separable
state. The system becomes entangled when the interaction energy of spins with
the magnetic field are of the order of the dipolar interaction energy. Then,
with increasing magnetic field the spin state tends to separable one. At a
small dipolar coupling constant $\left(  d\ll1\right)  $ from the exact
analytical solution (9) we obtain the following expression for the
concurrency
\begin{equation}
C=\max\left\{  0,-\frac{1}{2\cosh^{2}\frac{\beta}{2}}\right\}  \tag{12}%
\end{equation}
Therefore, at these conditions the states of the system are always separable,
$C=0$. Entanglement appears in the course of increasing the dipolar coupling
constant. To distinguish an entangled state from separable ones, it is
important to determine an entanglement witness (EW) applicable to
the\emph{\ }considered quantum system \cite{Horodecki1996,Terhal2000}. The
determination of EW is one of the main problems of the experimental study of
the entangled states. Internal energy \cite{Wang2002}, magnetic susceptibility
\cite{Wiesniak2005}, magnetization \cite{Brukner2004,FurmanQIP2009} , and
intensity of MQ coherences \cite{Fel'dman2008,FurmanPRA2009,FurmanQIP2009b}
were used as EW in different quantum systems. With the aim to obtain the
correlation between the nuclear magnetization and concurrence, using (4) and
the definition of nuclear magnetization $M_{z}=Tr\left(  \rho I_{z}\right)  $,
we obtain the exact expression for magnetization as a function of parameters
$\beta$ and $d$ at $\theta_{12}=\frac{\pi}{2}$ and $\varphi_{12}=0$: \bigskip%
\begin{equation}
M_{z}=\frac{-4\beta\sinh\frac{\sqrt{16\beta^{2}+9d^{2}}}{4}}{\left(
16\beta^{2}+9d^{2}\right)  \left(  \cosh\frac{\sqrt{16\beta^{2}+9d^{2}}}%
{4}+e^{\frac{d}{2}}\cosh\frac{d}{4}\right)  } \tag{13}%
\end{equation}
\ As example, at $d=3$, the relation between the concurrence and magnetization
can be fitted by $C=-0.71(M+0.26)$ (Figure 5). \ 

Concurrence, the measure of entanglement between the states of the two spins,
depends on the orientation of the magnetic field relative to vector $\vec{r}$
connecting the nuclei. At $\theta=0$ and $\pi$ the states are separable and
the concurrence reaches its maximum at the case of $\theta=\frac{\pi}{2}$
(Fig. 1). This effect can open a way to manipulate with the spin state by a
rotation of the magnetic field or a sample.

In conclusion, investigation of entanglement in a spin 1/2 system under the
thermodynamic equilibrium conditions showed that the entangled state can be
achieved by application of a low external field when the Zeeman interaction
energy is the order of or even less than the dipolar interaction one. It was
estimated that for magnetic field $H_{0}=$ $3$ G, the entangled state in a
two-spin system arises at temperature $T\lesssim0.33$
%TCIMACRO{\U{3bc}}%
%BeginExpansion
$\mu$%
%EndExpansion
K. The correlation between concurrence\emph{ }and nuclear magnetization is
considered and it was shown that concurrence is well fitted by a linear
dependence on the magnetization in the temperature and magnetic field range up
to a deviation of the magnetization from Curie's law $\left(  \beta
=3.32\text{, Fig. 5}\right)  $ .\emph{ }

\bigskip

\bigskip

\subsection{Figure Captions:}

Fig. 1 (Color online) Concurrence as a function of the parameter $\beta
=\omega_{0}/k_{B}T$ \ and magnetic field direction at $\varphi=0$ and $d=3$.

Fig. 2 The phase diagram. Line presents boundary between the entangled and
separated states in the plane\ $\beta-d$.

Fig. 3 (Color online) Concurrence as a function of the ratios of the magnetic
field strength $\left(  \omega_{0}\right)  $ and dipolar coupling constant
$(\frac{\gamma^{2}}{r_{12}^{3}})$ to $k_{B}T$.

Fig. 4 (Color online) Concurrence vs. magnetic field at $T=const$ (a) and vs.
temperature at $H=const$ (b) for various dipole interaction constants. (a):
black solid line - $d=0.5$; \ red dashed line - $d=2$ ; blue doted line -
$d=10.$ Magnetic field is given in units of $\frac{k_{B}T}{\gamma}$ . (b)
black solid line - \ $\frac{d}{\beta}=3$; red dashed line - $\frac{d}{\beta
}=5$ ; blue doted line - $\frac{d}{\beta}=10$. Temperature is given in units
of $\frac{\gamma^{2}}{r_{12}^{3}k_{B}}$.

Fig. 5 (Color online) Absolute value of magnetization (black solid line) and
concurrence (red dash line) as a function of $\beta=\frac{\omega_{0}}{k_{B}T}%
$. Fitting of the concurrence (blue dash-dot line) by $C=-0.71(M+0.26)$ at
$d=3$.


\begin{thebibliography}{99}                                                                                               %


\bibitem {Benenti books 2007}G. Benenti, G. Casati, and G. Strini,
\textit{Principles of Quantum Computation and Information}, Vol. I and II
(World Scientific, 2007).

\bibitem {Amico 2008 rev}Amico, L., Fazio, R., Osterloh, A. \& Vedral, V. Rev.
Mod. Phys. \textbf{80}, 517 (2008).

\bibitem {HorodeckiR 2009 rev}R. Horodecki, P. Horodecki, M. Horodecki, and K.
Horodecki, Rev. Mod. Phys. \textbf{81}, 865 (2009).

\bibitem {Bennett CH 1993}C. H.Bennett , G.Brassard , C.Crepeau , R.Jozsa ,
A.Peres , and W. K.Wootters , Phys. Rev. Lett. \textbf{70}, 1895 (1993).

\bibitem {Bennett CH 1984}C.H. Bennett and G.Brassard , Proceedings of IEEE
International Conference on Computers, Systems and Signal Processing,
Bangalore, India, pp. 175-179, December 1984.

\bibitem {Shor P 1994}Shor P., in Proceedings of 35th Annual Symposium on the
Foundations of Computer Science (IEEE Computer Society, Los Alamitos, CA,
1994), p. 124-134.

\bibitem {Cappellaro P 2005}P. Cappellaro , J. Emerson , N. Boulant , C.
Ramanathan , S. Lloyd , and D. G. Cory , Phys. Rev. Lett., \textbf{94}, 020502 (2005).

\bibitem {Roos2006}C.F.Roos, K.Kim, M.Riebe, R. Blatt, Nature \textbf{443},
316 (2006).

\bibitem {Bose S 2007}S.Bose , S. F.Huelga , D.Jonathan , P. L.Knight ,
M.Murao , M. B.Plenio and V.Vedral , \textit{Manipulation of Entangled States
for Quantum Information Processing, Quantum Communication, Computing, and
Measurement }, Edited by Kumar P. , D'Ariano G. M. and Hirota O., Kluwer
Academic / Plenum Publishers. New York, 2000

\bibitem {Fel'dman2007}S. I. Doronin, A. N. Pyrkov, and E. B. Fel'dman, JETP
Letters, \textbf{85}, 519 (2007).

\bibitem {Doronin2003}S. I. Doronin, Phys. Rev. A 68, 052306 (2003).

\bibitem {Fel'dman2008}E. B. Fel'dman and A. N. Pyrkov, JETP Lett.
\textbf{88}, 398 (2008).

\bibitem {Furman2008}G. B. Furman, V. M. Meerovich, and V. L. Sokolovsky,
Phys. Rev. A \textbf{78}, 042301 (2008).

\bibitem {Abragam 1982}A. Abragam and M. Goldman, \textit{Nuclear Magnetism:
Order and Disorder}, International Series of Monographs in Physics Clarendon,
Oxford, 1982.

\bibitem {Wootters1998}W. K. Wootters, Phys. Rev. Lett. \textbf{80}, 2245
(1998) .

\bibitem {Abragam 1974}M. Goldman, M. Chapellier, Vu Hoang Chau, and A.
Abragam , Phys. Rev. B \textbf{10}, 226 (1974).

\bibitem {Horodecki1996}M. Horodecki, P. Horodecki, and R. Horodecki, Phys.
Lett. A \textbf{223}, 1 (1996).

\bibitem {Terhal2000}B. M. Terhal, Phys. Lett. A \textbf{271}, 319 (2000).

\bibitem {Wang2002}X. Wang, Phys. Rev. A \textbf{66}, 034302 (2002).

\bibitem {Wiesniak2005}M. Wie\'{s}niak, V. Vedral, and C. Brukner, New J.
Phys. \textbf{7}, 258 (2005).

\bibitem {Brukner2004}C. Brukner and V. Vedral, e-print arXiv:quant-ph/0406040.

\bibitem {FurmanQIP2009}G. B. Furman, V. M. Meerovich, and V. L. Sokolovsky,
Quantum Inf. Process. \textbf{8}, 283 (2009).

\bibitem {FurmanPRA2009}G. B. Furman, V. M. Meerovich, and V. L. Sokolovsky,
Phys. Rev. A \textbf{80}, 032316 (2009).

\bibitem {FurmanQIP2009b}G. B. Furman, V. M. Meerovich, and V. L. Sokolovsky,
Quantum Inf Process, \textbf{8}, 379 (2009).
\end{thebibliography}
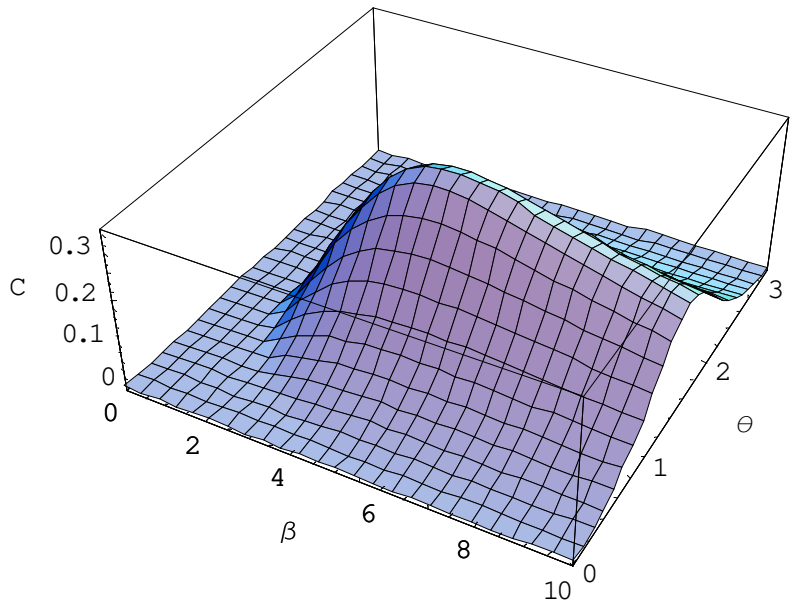
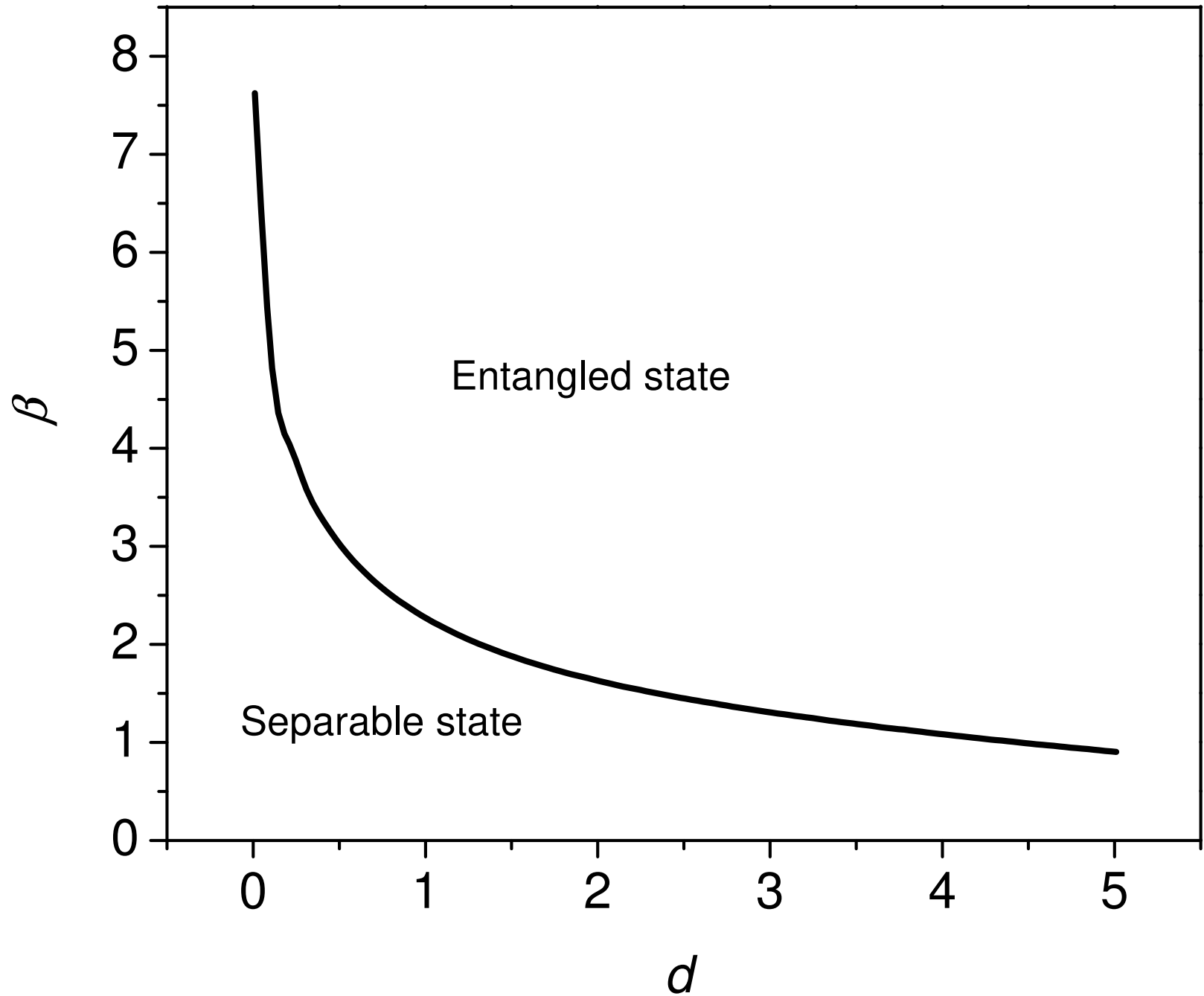
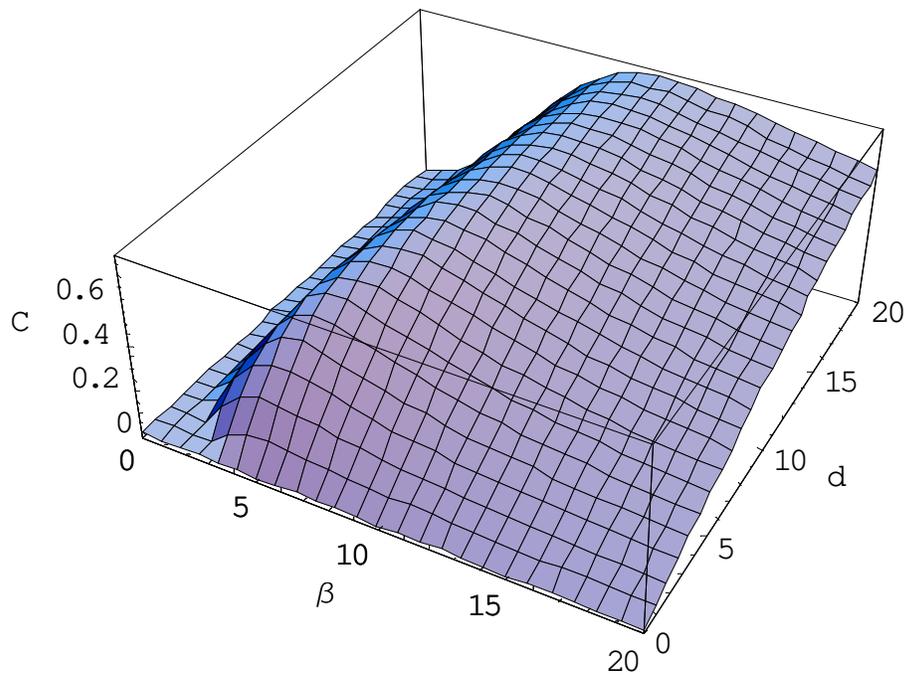
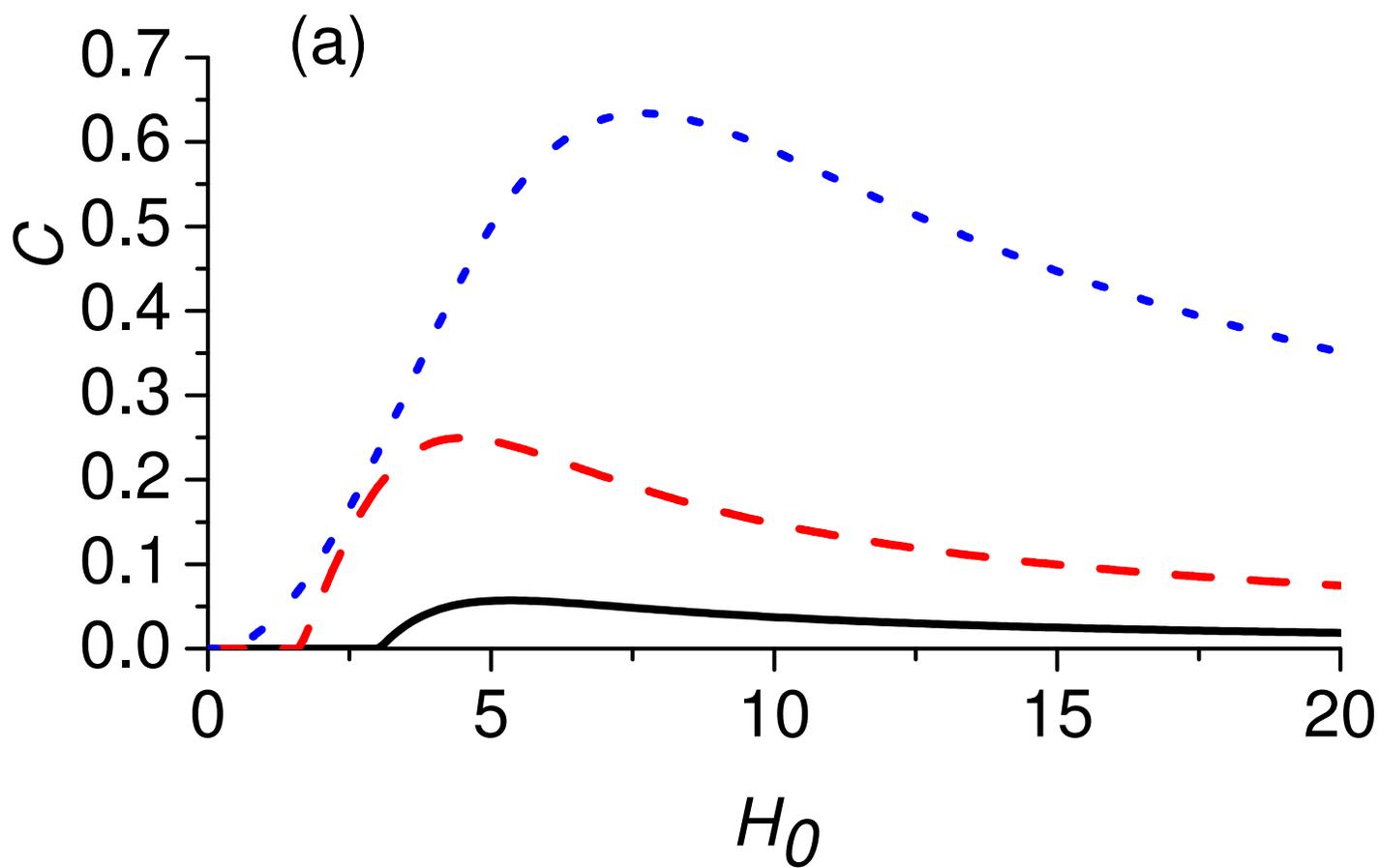
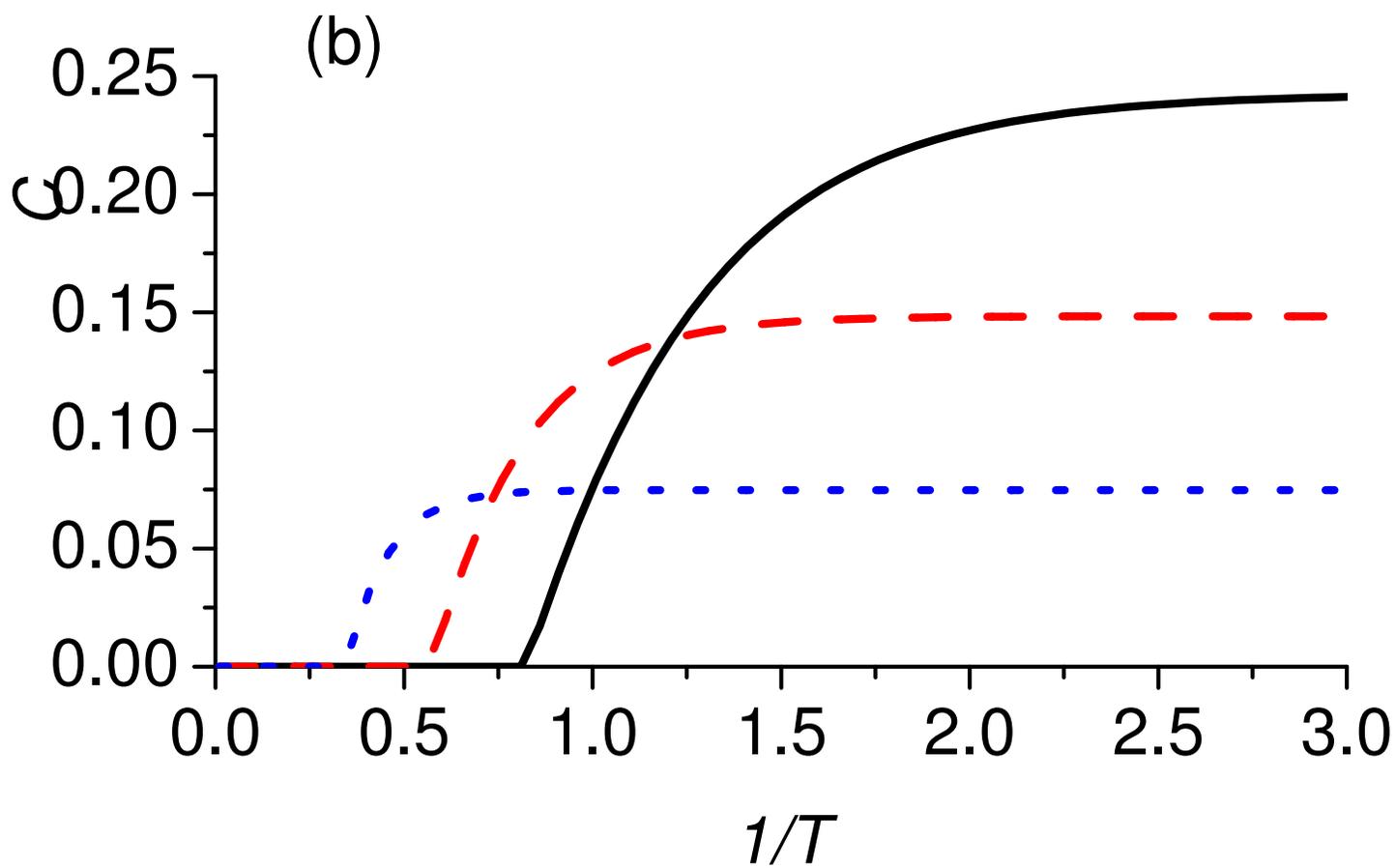
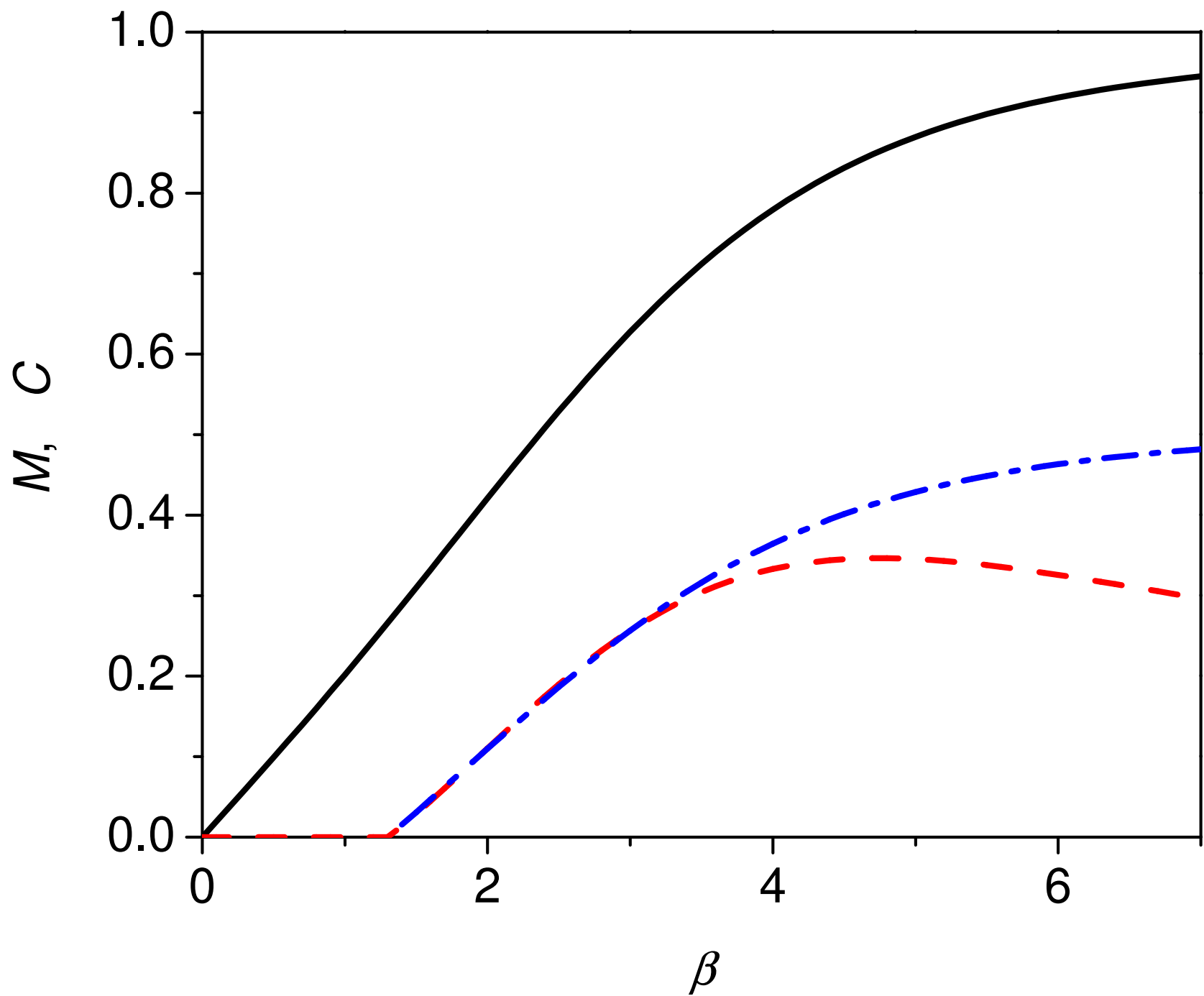
\end{document}